\documentclass[preprint,amsmath,amssymb]{revtex4}
\usepackage{graphicx}
\usepackage{dcolumn}
\usepackage{bm}
\usepackage{pstricks}
\usepackage{epsfig}
\usepackage{pst-grad}
\usepackage{pst-plot}
\begin{document}
\title{Fermi coordinates and modified Franklin transformation : A comparative study on rotational phenomena}
\author{ M.~Nouri-Zonoz  \footnote{Electronic address:~nouri@khayam.ut.ac.ir, corresponding author}
 and  H.~Ramezani-Aval  \footnote{Electronic address:~hramezania@ut.ac.ir}}
\address {Department of Physics, University of Tehran, North Karegar Ave., Tehran 14395-547, Iran.}
\begin{abstract}
Employing  a relativistic rotational transformation to study and analyze rotational phenomena, instead of the  rotational transformations based on consecutive Lorentz transformations and Fermi coordinates, leads to different predictions. In this article, after a comparative study between Fermi metric of a uniformly rotating eccentric observer and the spacetime metric in the same observer's frame obtained through the modified Franklin transformation, we consider  rotational phenomena including transverse Doppler effect and Sagnac effect in both formalisms and compare their predictions. We also discuss length measurements in the two formalisms.
\end{abstract}
\maketitle
\section{Introduction} \label{sec:introduction}
Rotation and rotating phenomena have always puzzled people and looking at the history of the theory of relativity (both special and general) it seems that rotating observers and their spatio-temporal measurements had a key role in forming Einstein's thoughts on the relation between non-inertial observers/frames and the gravitational field \cite{Ein}, encoded in the equivalence principle and the geometrical formulation of GR. Specifically the problem of a rigidly rotating disk and its spatial geometry seems to be one of the main elements in the development of general relativity, leading  to Einstein field equations in 1915 \cite{Stach}.
A general feature in studying rotational phenomena is the coordinate transformation between inertial (non-rotating) and rotating observers. This is not only a matter of convenience in using the right coordinates but also a matter of getting the most plausible interpretation for the measurements made by different observers and their relations. It is expected that the relation between the  spatio-temporal measurements in the two frames (rotating and non-rotating) should be different if one employs different kinematical transformations between them. Different kinematical transformations are also expected to result in spatially different flat spacetime metrics in the rotating observer's frame. The common practice in treating rotational phenomena is the employment of the so-called Galilean rotational transformation (GRT) between the rotating and non-rotating frames. Noting that in a rotating frame there are both inertial (centric) and non-inertial (eccentric) observers, one should be cautious with the restrictions in applying GRT which is only applicable to the former \cite{MFT}. By the same token, its application to eccentric observers is also questionable on the grounds that for relativistic rotational velocities one requires a relativistic rotational transformation (RRT), very much in the same way as Lorentz transformation replacing the Galilean transformation among inertial frames at relativistic velocities. Proposals for an RRT date as far as back to the 1920s and the introduction of the first relativistic rotational transformation by Philip Franklin \cite{Franklin}. On the other hand the usual approach to study and analyze physical phenomena in accelerated frames (of which the uniformly rotating frame is a special case), in flat and curved spacetimes, employs the so-called hypothesis of locality \cite{Mashhoon}. This hypothesis asserts that accelerated observers are instantaneously equivalent to hypothetical inertial observers having the velocity of the accelerated observer at each moment on its worldline. Therefore to find the coordinate transformation between two different positions of the accelerated observer on its  worldline, one uses Lorentz transformations between the corresponding hypothetical inertial frames and a reference inertial frame. For example, consider an observer at a given radius on a uniformly rotating platform. The coordinates that this observer assigns to an event at its two different rotational positions $A$ and $B$ around the origin $O$ is found by a Lorentz transformation from the hypothetical inertial frame at $A$ to the central inertial frame $O$, followed by a Lorentz transformation from $O$ to the hypothetical inertial frame at $B$. This is the same process which leads to the so-called Thomas precession.\\
Therefore to study physical phenomena in  a uniformly rotating eccentric observer's frame  and relate the measurements in such a frame to those in an inertial frame, one can either use an RRT or employ the hypothesis of locality and the resultant coordinate transformation.\\
To emphasize once more the fundamental difference between the two approaches, we note that at the heart of the two approaches lie two different kinematical transformations. In the first approach,  one uses an RRT, such as the modified Franklin transformation employed in the present study, which is fundamentally different from LT. In the second approach the hypothesis of locality is employed, which is based on  consecutive Lorentz transformations among hypothetical inertial observers which are instantaneously equivalent to the accelerated one at each moment on its world line.\\
It should be noted that it is the implicit application of the same hypothesis which leads to the so-called Fermi coordinates and Fermi metric that an accelerated, spatially rotating observer would assign to his/her reference frame \cite{mtw}. It is interesting to find that Fermi coordinates were introduced in exactly the same year as the Franklin transformation was introduced \cite{Fermi}. Our main goal here is to study these two approaches and compare their predictions for the well known rotational phenomena as measured by non-inertial rotating observers. The plan of the paper is as follows. In the next section we introduce the Fermi coordinates of an accelerated spinning observer and in section III the same coordinates are used to find the Fermi metric in a rotating frame. In section IV modified Franklin transformation as an RRT, relating coordinates of events in  rotating  and inertial (non-rotating) frames, is introduced and the spacetime metric based on this transformation is given in the rotating observer's frame. In section V both transverse Doppler effect (TDE) and Sagnac effect, in a rotating observer's frame, are studied comparatively by applying these two different approaches. In the same section the relation between length measurements in rotating and inertial frames will be discussed in both approaches. In what follows Roman indices run from $1$ to $3$ while Greek ones run from $0$ to $3$ and our metric signature is $(-,+,+,+)$.
\section{Accelerated, rotating Observers and Fermi Coordinates } \label{Math}
As pointed out in \cite{mtw} {\it `` it is very easy to put together the words ''the coordinate system of an accelerated observer`` but it is much harder to find a concept these words might refer to''} and it gets even harder, at least conceptually, if one wishes to extend it to {\it accelerated observers} in curved spacetimes. One could assign a coordinate system to an accelerated spinning observer, who carries an orthonormal tetrad, both in flat and curved spacetimes. Although the formalism, to first order in the spatial coordinates, looks the same in flat and curved spacetimes, it should be noted that the main difference is in the size of the region in which such a coordinate system is applicable. In the case of flat spacetime it could be applied to a region within a finite distance from the observer, but in the case of curved spacetime it is restricted to an infinitesimal neighborhood of the tetrad's origin on the observer's world line over which the curvature is not felt \footnote{For general restrictions on possible extended reference frames for accelerated observers refer to \cite{Marzlin}.}.
In this approach setting the origin of the accelerated observer's frame (S) on the observer's world line, the orthonormal
tetrad $e^{\mu}(\tau)$ (consisting of one timelike and three spacelike vectors) carried by the observer is Fermi-walker transported \footnote{For a historical account on the Fermi coordinates and Fermi-Walker transport refer to \cite{Bini}.} and the coordinate transformation between an inertial (Laboratory) observer and the accelerated observer is given by \cite{mtw}
\begin{eqnarray}\label{1}
{x^{\prime}}^{\mu}=x^{k}[e_k(\tau)]^{\mu}+Z^{\mu}(\tau)\;\;\;\; k=1,2,3
\end{eqnarray}
in which $\tau$ is the observer's proper time and  $Z^{\mu}(\tau)$ is its worldline
relative to an inertial frame. ${x^{\prime}}^\mu$ and $x^{k}$ are
the coordinates assigned to an event in the inertial frame ($S^{\prime}$) and
in the accelerated observer's local frame (S) respectively. Now if the accelerated observer has also a spatial rotation, its tetrad endowed with an angular velocity, is not Fermi-Walker transported along his worldline but is transported according to the following rule
\begin{eqnarray}\label{2}
\frac{d [{e}_{\alpha}]^{\mu}}{d\tau}=-{\Omega^{\mu\nu}}{[{e}_{\alpha}]_\nu}
\end{eqnarray}
where  $\alpha$ is the tetrad (Lorentz) index and
\begin{eqnarray}\label{3}
\Omega^{\mu\nu}=a^\mu u^\nu-a^\nu u^\mu + u_\alpha \Omega{_\beta}
\epsilon^{\alpha\beta\mu\nu}
\end{eqnarray}
is composed of two parts, the first part made out of  ${u}^\mu$ and ${a}^\mu$ (i.e. the 4-velocity and 4-acceleration of the observer), indicates the Fermi-Walker part while the second part (last term in (\ref{3})), which includes the observer's 4-rotation $\Omega^\mu$, denotes the spatial rotation. The first order expression for the metric near the observer's world line is given by the so-called Fermi metric \cite{Fermi},
\begin{eqnarray}\label{4}
ds^2=-(1+2a''_{l}x^{l}){dx^{0}}^2-2(\epsilon_{jkl}x^{k}\Omega''^{l})dx^{0}dx^{j}+\delta_{ij}dx^{i}dx^{j}
\end{eqnarray}
in which ${\bf a}^{\prime\prime}$ and ${\bf \Omega}^{\prime\prime}$
are the observer's acceleration and (spin) rotation measured in a comoving
inertial frame $(S^{\prime\prime})$ whose velocity is momentarily
the same as that of the accelerating observer. In \cite{WTN} by
extending this method to the second order in $x^i$, the metric in 
an accelerated spinning frame in curved spacetime, is derived as follows,
\begin{eqnarray}\label{5}
ds^2=-{dx^{0}}^2[1+2a''_{j}x^{j}+(a''_{l}x^{l})^2+(\Omega''_{l}x^{l})^2-\Omega^{''2}x_{l}x^{l}+R_{0l0m}x^{l}x^{m}]\nonumber
\\
+2dx^{0}dx^{i}(\epsilon_{ijk}\Omega''^{j}x^{k}-\frac{2}{3}R_{0lim}x^{l}x^{m})+dx^{i}dx^{j}(\delta_{ij}-\frac{1}{3}R_{iljm}x^{l}x^{m})
\end{eqnarray}
which in flat spacetime reduces to
\begin{eqnarray}\label{6}
ds^2=-{dx^{0}}^2[1+2a''_{l}x^{l}+(a''_{l}x^{l})^2+(\Omega''_{l}x^{l})^2-\Omega''^2x_{l}x^{l}]\nonumber
\\
+2dx^{0}dx^{i}(\epsilon_{ijk}\Omega''^{j}x^{k})+dx^{i}dx^{j}\delta_{ij}.
\end{eqnarray}
In \cite{Nelson}, looking for a  generalization of the Lorentz transformation to the case of accelerated rotating observers (with a
time-dependent velocity), a nonlinear coordinate transformation was introduced which, not only incorporates the Thomas precession, but also leads to the above spacetime metric exactly. Two of the main properties of metric (\ref{6}) are as follows \cite{Mashhoon1,Nikolic},\\
I-In the absence of any linear acceleration ($a=0$) this metric (in the Cartesian coordinates) reduces to
\begin{eqnarray}\label{7}
ds^2=-[c^2 -(x^2+y^2)\Omega^2]{dt}^2+{dx}^2+{dy}^2+{dz}^2-2y\Omega{dxdt}+2x\Omega{dydt}
\end{eqnarray}
which is the Galilean rotational metric assigned to the flat spacetime by an
observer at $r=0$ with constant angular velocity $\Omega$. The same metric in cylindrical coordinates $(t,r,z,\phi)$ is given by \cite{Landau}
\begin{eqnarray}\label{7-1}
ds^2 = -(c^2-r^2\Omega^2){dt}^2 + dr^2 + {dz}^2 + r^2 d\phi^2 + 2 \Omega r^2{d\phi}dt 
\end{eqnarray}
II- In the absence of any spatial rotation ($\Omega=0$), as expected, this metric reduces to the Rindler metric,
\begin{eqnarray}\label{8}
ds^2=-{dx^{0}}^2[1+2a''_{j}x^{j}+(a''^{l}x^{l})^2]+dx^{i}dx^{j}\delta_{ij}
\end{eqnarray}
which is the metric of the flat spacetime in the proper frame of a uniformly accelerated observer.
Obviously in case both $a$ and $\Omega$ are zero, the observer would be a freely falling one.
\section{Uniformly rotating observers in Fermi coordinates}\label{URO}
Having discussed the general case of an accelerated rotating observer and the corresponding  spacetime metric in the previous section, here we are interested in the particular case of an observer who moves on a circular path such as the one fixed on a uniformly rotating disk. 
Both in \cite{Mashhoon1} (based on the hypothesis of locality) and in \cite{Nikolic} (based on the formulation introduced in \cite{Nelson}), 
coordinate transformation between such an
observer and an inertial (laboratory) observer has been introduced. In their setup the origin of the rotating frame is on the
circular path and oriented such that its $y$-axis is always tangent to the circular path. Also at $t=0$ the axes of the inertial and  rotating frames are taken to be parallel (Fig. 1). 
\begin{figure}
\begin{center}
\includegraphics[angle=0,scale=0.7]{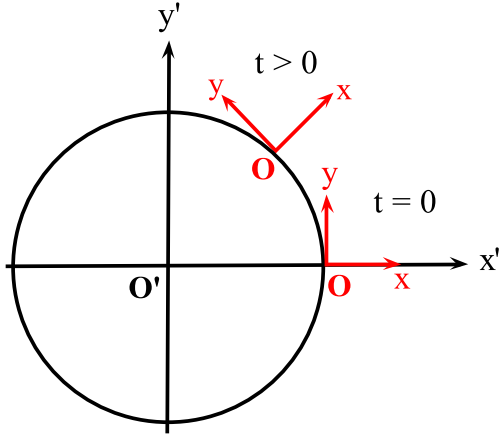}
\caption{Local coordinates of a rotating observer on a circular path with axes $x$ and $y$ which are always along the 
radius of the path and tangent to it respectively.}
\end{center}
\end{figure}
Therefore as in the case of Fermi metric for an accelerated, rotating observer, the origin of the rotating observer's frame is  on the observer's
world line. Furthermore, the planar axes of the rotating observer's frame are always along the 
radius of the path and tangent to it (Fig. 1). In this way the angular frequency of the observer's rotation about the inertial observer's frame ($O^\prime$) would  be equal to the observer's spinning frequency about the origin of his local frame ($O$). This setup is automatically satisfied in the case of a fixed observer on a uniformly rotating disk with initially aligned axes. It may seem that this special
case reduces the generality of the problem, but it should be noted that such construction of frames is important
as they are the ones which are related to the real experimental setups such as in the case of observers/detectors on
the circumference of a uniformly rotating disk. In
\cite{Mashhoon1} after introducing the corresponding tetrad  for such an
observer and using the same  method employed in obtaining  the Fermi metric in \cite{mtw}, the following coordinate transformation was introduced
between an inertial (primed) frame and a frame uniformly rotating (unprimed) about it with angular velocity $\Omega$, 
\footnote{To prevent any confusion it should be noted that up to now 
$\Omega$ represented the spinning angular velocity of a frame but hereafter it will represent its orbital angular velocity. It should also be noted that in the setup given in Fig. $1$ (e.g. for a frame fixed on the rim of a rotating disk) the two angular velocities have the same value.}
\begin{eqnarray}\label{9}
ct=\gamma^{-1}(ct^{\prime}-\beta\gamma
y)~~~x={x^{\prime}}\cos(\gamma\Omega
t)+{y^{\prime}}\sin(\gamma\Omega t)-R \nonumber 
\\
y=\gamma^{-1}[- x^{\prime}\sin(\gamma\Omega
t)+y^{\prime}\cos(\gamma\Omega
t)]~~~~,~~~~z=z^{\prime}
\end{eqnarray}
in which $\gamma=(1-\beta^2)^{-1/2}$, $\beta=\frac{R\Omega}{c}$ and $R$ is
the radius of the circular path traveresed by an eccentric observer, as measured in an inertial observer's frame. Note that at $t=0$ we have $x=x^\prime -R$ and $y=\gamma^{-1} y^\prime$ which are in accordance with the alignment of the local coordinates at that time as shown in Fig. 1. By the above transformation the circular path  ${x^\prime}^2 + {y^\prime}^2=R^2 $ in the inertial frame is an ellipse $(x+R)^2 + \gamma^2 y^2 = R^2$ with a contracted circumference (refer to section V) in the rotating observer's frame \cite{Mashhoon2}.
Using the general Lorentz transformation introduced in \cite{Nelson} or the orthonormal tetrad frame of a rotating observer, the inverse of the above coordinate transformations is given in \cite{Nikolic,Mashhoon2} as follows,
\begin{eqnarray}\label{12}
\left(
  \begin{array}{c}
    ct' \\
    x' \\
    y' \\
    z' \\
  \end{array}
  \right)={\left(
                \begin{array}{cccc}
                  \gamma & 0 & \gamma\Omega R/{c} & 0 \\
                  0 & \cos(\gamma\Omega{t}) & -\gamma\sin(\gamma\Omega{t}) & 0 \\
                  0 & \sin(\gamma\Omega{t}) & \gamma\cos(\gamma\Omega{t}) & 0 \\
                  0 & 0 & 0 & 1 \\
                \end{array}
              \right)}\left(
                        \begin{array}{c}
                          ct \\
                          x \\
                          y \\
                          z \\
                        \end{array}
                      \right)
              +R \left(
                         \begin{array}{c}
                           0 \\
                           \cos(\gamma\Omega{t}) \\
                           \sin(\gamma\Omega{t}) \\
                           0 \\
                         \end{array}
                       \right)
\end{eqnarray}
Using the differential form of the above coordinate transformations and substituting them into the Minkowski metric of the inertial observer, 
the line element in the rotating  frame is given by,
\begin{eqnarray}\label{10}
ds^2=-\gamma^{2}[c^2-(R+x)^{2}\Omega^2-\Omega^2 
{y}^{2}]{dt}^2+{dx}^2+{dy}^2+{dz}^2-2y\Omega{dxdt}+2x\Omega{dydt}
\end{eqnarray}
Comparison with the general Fermi metric (\ref{6}) reveals the following  3-vectors
\cite{Nikolic}
\begin{eqnarray}\label{11}
{{a}^{''}}^l=(-\gamma^{2}R\Omega^2,0,0) ~~~,~~~
{{\Omega}''}^l=(0,0,\gamma^{2}\Omega)
\end{eqnarray}
as the observer's acceleration and angular velocity measured by the comoving inertial frame. 
In other words for an observer with the above 3-acceleration and 3-angular velocity (\ref{6}) concludes (\ref{10}).
Since the acceleration  is proportional to $R$, it can be seen that at $R=0$ ( with $x,y \neq 0$) the above metric
(\ref{10}) reduces to the metric (\ref{7}) which could be obtained by applying the GRT to the flat spacetime metric 
of an inertial observer in Carteian coordinates.
This is expected as the transformation \eqref{9} itself reduces to GRT at $R=0$. On the other hand setting $x=y=0$ (with $R\neq 0$) in \eqref{10}
, i.e. at the position of the rotating observer in its local frame, it reduces to the Cartesian flat spacetime metric in accordance with characteristics of the Fermi metric. \\
The limitations in length measurements by a uniformly
rotating observer in this construction are discussed in \cite{Mashhoon2}.
Although existence of a rotating  disk is not considered explicitly in the above construction of the Fermi metric, a
uniformly rotating observer as introduced here (with equal spinning and
orbiting angular frequencies) could be realized in an experimental setup with a detector at a nonzero radius on a uniformly rotating  disk.
As we will discuss later, this is basically the experimental setup used to investigate transverse Doppler effect as a rotational phenomenon.
\section{Uniformly otating eccentric observers and modified Franklin transformation}
In \cite{MFT}, looking for a consistent relativistic rotational transformation between an inertial observer (frame $S^\prime$)
and an observer at a nonzero radius (eccentric observer) on a uniformly rotating disk (frame S) \footnote{Since here we are interested in the quantities from the rotating observer's perspective, the assignment of primed and unprimed frames are opposite to that in \cite{MFT}.}, the following modification of 
the so-called Franklin transformation (in cylindrical coordinates) was introduced,
\begin{eqnarray}\label{13}
t =  \cosh (\Omega R/c)t^{\prime} - \frac{R}{c}  \sinh (\Omega
R/c)\phi^{\prime} \;\;\; ; \;\;\; r = r^{\prime} \cr
\phi = \cosh (\Omega R/c) \phi^{\prime} -  \frac{c}{R} \sinh
(\Omega R/c) t^{\prime} \;\;\; ; \;\;\; z = z^{\prime},
\end{eqnarray}
in which $\Omega$ is the uniform angular velocity of the disk and $R$ is the radial position of the observer on the disk. Note that the origin of the rotating frame $S$ is chosen to be at the center of the rotating disk so that both the inertial and rotating frames assign the same radial coordinate to an event (Fig. 2). The corresponding metric in the rotating observer's frame is given by ($\beta =\frac{\Omega R}{c} $),
\begin{eqnarray}\label{14}
ds^2 = -c^2\cosh^2\beta(1- \frac{{r}^2}{R^2}\tanh^2\beta) {dt}^2 +
{dr}^2 + {r}^2 \cosh^2\beta  \cr (1-
\frac{R^2}{{r}^2}\tanh^2\beta)d{\phi}^2 -2cR \sinh \beta \cosh \beta
(1- \frac{{r}^2}{R^2})dt d\phi  + {dz}^2.
\end{eqnarray}
As in the case of Franklin transformation, this is the flat spacetime metric with non-Euclidean spatial sector. But contrary to the spacetime metric obtained in the rotating observer's frame through Franklin transformation, it reduces to the spacetime metric obtained via GRT in the limit $\beta \ll 1$, i.e. close to the rotation axis \cite{MFT} where the rotational velocity is non-relativistic. Also as in the case of Fermi metric, at the position of the observer, i.e., $r=R$,
this metric reduces to that of spatially Euclidean Minkowski metric in cylindrical coordinates. To compare the above metric for a rotating observer with that obtained for the same observer in Fermi coordinates, we rewrite it in the Cartesian coordinates as follows,
\begin{eqnarray}\label{15}
ds^2 = -c^2[\cosh^2\beta - \frac{x^2+y^2}{R^2}\sinh^2\beta] {dt}^2 +
[\frac{x^2}{r^2}+(\cosh^2\beta -
\frac{R^2}{r^2}\sinh^2\beta)\frac{y^2}{r^2}]{dx}^2\nonumber
\\
+[\frac{y^2}{r^2}+(\cosh^2\beta -
\frac{R^2}{r^2}\sinh^2\beta)\frac{x^2}{r^2}]{dy}^2 -
\sinh^2\beta(1-\frac{R^2}{r^2})\frac{xy}{r^2}{dxdy}\nonumber
\\
+2R \sinh \beta \cosh \beta (\frac{1}{r^2}- \frac{1}{R^2}){y}dtdx
-2R \sinh \beta \cosh \beta (\frac{1}{r^2}- \frac{1}{R^2})x dtdy +
{dz}^2.
\end{eqnarray}
Since both metrics at the position of the observer reduce to the
spatially Euclidean flat spacetime metric, to compare them, the above
line element is expanded around the position of the observer at 
$(x_0=R, y_0=0)$ (Fig. 2), leading to the following nonzero components of the metric:
\begin{eqnarray}\label{16}
g_{00}=-1+\frac{2\sinh^2\beta}{R}\xi_{1}+\frac{\sinh^2\beta}{R^2}({\xi_{1}}^2+{\xi_{2}}^2)~~~,~~~g_{02}=\frac{2\sinh
\beta \cosh \beta}{R}\xi_{1} \nonumber \\
g_{11}=1~~~,~~~g_{22}=1+\frac{-2\sinh^2
\beta}{R}\xi_{1}~~~,~~~g_{33}=1,
\end{eqnarray}
in which $\xi_i = (\xi_1,\xi_2)$ represents a small (Cartesian) displacement from the position of the rotating observer at $O$. 
By expanding $\sinh\beta$ and $\cosh\beta$ in (\ref{16}) and also 
$\gamma$ in (\ref{10}) to the second order in  $\beta = \frac{R \Omega}{c}$ it can easily be seen that the time-time
components $g_{00}$ in the two metrics agree (identifying $ \xi_i$ with $x_i=(x,y)$ in (\ref{10})) to the same order in  $\beta$.
Consequently it is expected that the rotational effects originating from the time-time component of the metric
in both approaches lead to the same predictions up to the second order in  $\beta = \frac{R \Omega}{c}$.
This assertion will be examined in the next section where we comparatively study rotational  phenomena
from the perspective of inertial (non-rotating) observers and eccentric (non-inertial) observers on a rotating platform, employing the 
two approaches based on MFT and hypothesis of locality.
\begin{figure}
\begin{center}
\includegraphics[angle=0,scale=0.7]{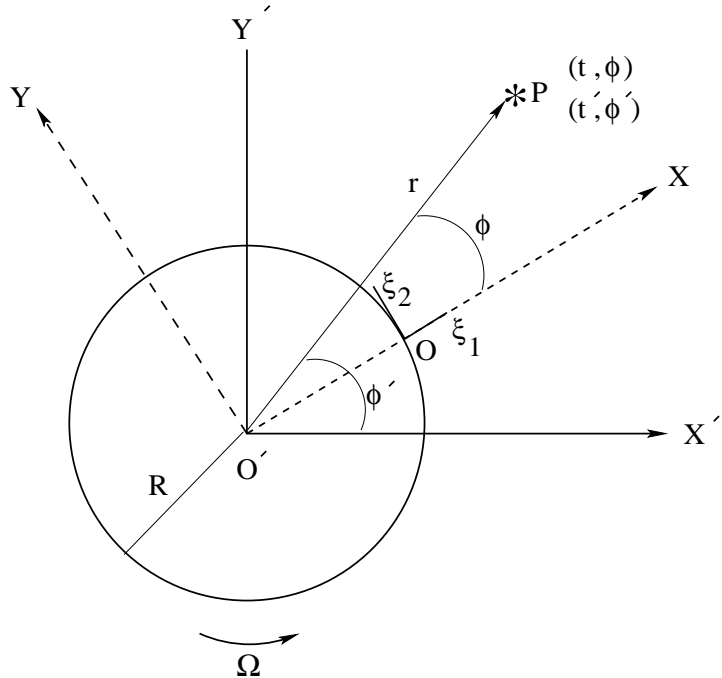}
\caption{Inertial frame $(X', Y')$ and the frame $(X, Y)$ of a centrally rotating observer which are always parallel to those used by the  observer $O$ at the rim of a uniformly rotating disk with radius $R$.
In cylindrical coordinates, an event $P$ has temporal and angular coordinates $(t, \phi)$ and $(t^\prime, \phi^\prime)$ in rotating and inertial frames, respectively.}
\end{center}
\end{figure}
\section{Application to rotational phenomena}
In the last two sections we have introduced two different kinematical transformations between an inertial frame and a rotating non-inertial frame. They were based on hypothesis of locality (consecutive Lorentz transformations) and an exact relativistic rotational transformation (modified Franklin transformation) respectively. Consequently they led to two different spatially non-Euclidean flat spacetime metrics in the rotating frame, with the same time-time component to the second order in $\beta = {\frac{\Omega R}{c}}$. To compare their predictions, here we are going to study the application of these transformations to rotational phenomena. In each case we  employ the transformations introduced in the two formalisms to obtain the relation between the physical quantities in the inertial (non-rotating) and rotating frames. These are then expanded in terms of the parameter $\beta = \frac{\Omega R}{c} $ to see how the results differ in the two formalisms  from those obtained through the classical (non-relativistic) treatment in which GRT or its equivalent metric is used. In other words the comparison is made between the results obtained in the two relativistic approaches at the classical limit $ {\Omega R} \ll {c}$. It should be emphasized again that although in the formalism based on Fermi coordinates and/or hypothesis of locality, a rotating disk or platform is not mentioned explicitly, but the authors employ what they call ``{\it a rotating measuring device}'' and the measurements are done in the comoving frame of this device \cite{Mashhoon2}, which in principle could be mounted on a rotating platform at a nonzero radius. In what follows the results of the measurements of such an observer/device are compared with those based on MFT when we set $r=R$, that is, in the comoving frame of an eccentric observer fixed at radial distance $R$ from the rotation axis on a rotating platform. In other words one should note that, in general, the radial coordinate of an event  $r= \sqrt{{x^\prime}^2 + {y^\prime}^2}$ could be different from $R$ the position of an observer/device (Fig. 2). This distinction makes the limit $R \rightarrow 0$ meaningful in the kinematical interpretation of either of the transformations \cite{MFT}.  
\subsection{Transverse Doppler Effect}
Transverse Doppler effect is a direct consequence of the time dilation in special relativity. In the simplest setup, an observer moving  on a rotating disk will measure the frequency of a light signal sent from a centrally located source. So to examine the effect of employment of a relativistic rotational transformation or consecutive Lorentz transformations based on the hypothesis of locality, one should examine the relation between the time intervals in the inertial and rotating frames. In this regard, the relation between the time intervals in an inertial observer's frame and a uniformly rotating one in the formalism based on the Fermi coordinates is given by the first equation in (\ref{9}). Looking at that equation it is clear that the transverse Doppler effect in this approach is the same as what one gets in the usual special relativistic treatment \cite{Rindler}. If the source
and receiver are both on the rotating disk at radii $R_1$ and $R_2$
respectively, the ratio of the emitted frequency to that received is given by
\begin{eqnarray}\label{25}
\frac{\nu_1}{\nu_2}=\frac{\gamma_1}{\gamma_2}\approx 1+\frac{\Omega^2}{2c^2}(R_1^{2}-R_2^{2})+\frac{\Omega^4}{c^4}(\frac{3}{8}R_1^{4}-\frac{1}{8}R_2^{4}-\frac{1}{4}R_1^{2}R_2^{2}).
\end{eqnarray}
On the other hand, using the same setup in the context of the MFT, by the time transformation in (\ref{13}) one arrives at the following result:
\begin{eqnarray}\label{26}
\frac{\nu_1}{\nu_2}=\frac{\cosh\beta_1}{\cosh\beta_2} \approx 1+\frac{\Omega^2}{2c^2}(R_1^{2}-R_2^{2})+\frac{\Omega^4}{c^4}(\frac{1}{24}R_1^{4}-\frac{5}{24}R_2^{4}-\frac{1}{4}R_1^{2}R_2^{2}),
\end{eqnarray}
which  differs from the  previous result in the third term which is of fourth order in $\beta = \frac{\Omega R}{c}$. This verifies our expectation on the rotational effects related to the time-time component of the metric in the rotating frame in the two formalisms. The same result could also be obtained by using the corresponding metric (\ref{14}) in the rotating frame and noting that the proper time at the position of an observer at  $r=R_1$ is given by
\begin{eqnarray}\label{26-0}
{d\tau_1}^2 = \cosh^{-2}\beta_1 {d t^\prime}^2,
\end{eqnarray}
in which $\beta_1 =\frac{\Omega R_1}{c}$ and $t^\prime$ is the coordinate (proper) time in the inertial observer's frame. So for two observers fixed at two different radii $r=R_1$ and $r=R_2$, the frequencies measured in terms of their proper times are related by
\begin{eqnarray}\label{26-00}
\frac{\nu_1}{\nu_2} \equiv \frac{d\tau_2}{d\tau_1} = \frac{\cosh\beta_1}{\cosh\beta_2}.
\end{eqnarray}
In either of the relations (\ref{25}) and (\ref{26}), one could set $R_1 =0$  and $R_2 = R_0$ (with $R_0$ the radius of the rotating disk) to find the frequency ratio for the case in which the source and receiver are at the center and rim of the disk respectively, so that (\ref{25}) and (\ref{26}) reduce to
\begin{eqnarray}\label{25-1} 
\frac{\nu_1}{\nu_2}=\sqrt{1 -\frac{{R_0}^2 \Omega^2}{c^2}} \approx 1 - \frac{\Omega^2}{2c^2}{R_0}^{2} -
\frac{1}{8}\frac{\Omega^4}{c^4}R_0^{4},
\end{eqnarray}
and
\begin{eqnarray}\label{26-1}
\frac{\nu_1}{\nu_2}= \sqrt{1 -{\rm tanh}^2 \frac{R_0 \Omega}{c}} \approx 1 - \frac{\Omega^2}{2c^2}{R_0}^{2} -
\frac{5}{24}\frac{\Omega^4}{c^4}{R_0}^{4},
\end{eqnarray}
respectively. In this way one could observe that in the formalism based on MFT, the frequency ratio arises from the same relation as in the special relativistic case but now with the nonlinear velocity $v=c \tanh \beta$ replacing the classical relation $v=R_0\Omega$ \cite{MFT}. Indeed the above configuration of the source and receiver is the same as that in the original experimental setups in which the m{\"{o}}ssbauer effect was used to verify transverse Doppler effect \cite{Hay,Kundig}. For example, taking into account many side effects such as the stretching of the rotor, K{\"{u}}ndig finds his experimental results to be in agreement with the theoretical prediction  based on Lorentz transformation and linear velocity distribution $v= r\Omega$ (to the second order in $\beta$) within $1\%$ error. Since the theoretical predictions based on MFT agree with those based on Lorentz transformations up to the second order in $\beta$, one  needs to carry out the same experiment ($\Omega$= 35000rpm, $R_0 = 9.3 cm$) with a precision of at least  $1$ part in $10^{14}$ to find any deviations in the fourth order. For the treatment of the same effect using an alternative RRT refer to \cite{Hsu}.
\subsection{Sagnac Effect}
Perhaps one of the most famous rotational effects is the so-called Sagnac effect \cite{Sag}, in which an interferometer on a rotating platform measures the effect of rotation on the phases of counterrotating photon beams. 
For two such beams starting at the same point on a rotating platform with uniform angular velocity $\Omega$, the difference in their arrival time to the initial point, as measured in an {\it inertial} frame, is given by \cite{Post}:
\begin{eqnarray}\label{2600}
{\Delta t}^{\prime} = \frac{4\pi {R}^2 \Omega}{c^2(1-\frac{R^2 \Omega^2}{c^2})}, 
\end{eqnarray}
which, in the Galilean limit $\beta = \frac{\Omega R}{c} \ll 1$, expanded in terms of $\beta$ gives
\begin{eqnarray}\label{260}
{\Delta t}^{\prime} = 4\pi {R}^2 \frac{\Omega}{c^2}(1 + \beta^2 + {\cal O}(\beta^4)),
\end{eqnarray}
in which $R$ is the radius of the circular  path traversed by the two beams. This time difference leads to a phase shift 
$\delta \phi = \frac{2\pi c \Delta t}{\lambda}$ \cite{Post}. The same effect could also be analyzed classically from a rotating observer's point of view by using the fact that in such a frame the spacetime metric, although flat, is in a stationary form given by the Galilean transformed metric (\ref{7-1}) which has  a non-Euclidean spatial sector. It should  be noted that Obviously in this context the rotating observer is the centrally rotating one which is in principle an inertial observer \cite{MFT}. In the context of the $1+3$ (or threading) formulation of spacetime decomposition \cite{Lynd}, it could be shown that this non-Euclidean character is rooted in the cross term of the corresponding stationary metric, and consequently the synchronization along a closed path  will lead to the following time difference (desynchronization) on returning to the departure point \cite{Landau,MFT}:
\begin{eqnarray}\label{new1}
\Delta t^\prime = -\frac{1}{c} \oint \frac{g_{0\alpha}}{g_{00}} dx^\alpha.
\end{eqnarray} 
In the case of the metric (\ref{7-1}) and for a circular path of radius R  on a rotating disk, it reduces to 
\begin{eqnarray}\label{new11}
\Delta t_{\pm}^\prime = \pm \frac{1}{c^2}\oint \frac{\Omega R^2 d \phi}{1- \frac{\Omega^2 R^2}{c^2}} = \pm \frac{2\pi {R}^2 \Omega}{c^2(1-\frac{R^2 \Omega^2}{c^2})},
\end{eqnarray} 
in which the $\pm$ signs refer to the corotating and counterrotating paths. Obviously their difference leads to the same relation ({\ref{2600}). In this formulation the Sagnac effect and the so-called clock effect \cite{Coh} are treated as different manifestations (null and timelike) of desynchronization effect in axisymmetric stationary spacetimes. \\
Using a simple instantaneous Lorentz transformation, one can relate the above inertial time difference to that in the rest frame of the non-inertial observer at radius $R$, as follows \cite{Post}:
\begin{eqnarray}\label{new2}
{\Delta t} = \gamma^{-1}{{\Delta t}^{\prime}}, 
\end{eqnarray}
in which $\gamma = \frac{1}{\sqrt{1-\beta^2}}$. To study this rotational effect in the context of the formalism based on  MFT, we use the time transformation relation in (\ref{13}) between the time intervals of the two events corresponding to the arrival of the two counterrotating light beams (departing at the same time $t_0$ in frame S) to the same point on the rotating disk, 
\begin{eqnarray}\label{28}
{\Delta t} = (t_2 - t_0) - (t_1 - t_0)= t_2- t_1 = \cosh\beta(t^{\prime}_2-t^{\prime}_1) - \frac{R}{c}\sinh\beta(\varphi^{\prime}_2-\varphi^{\prime}_1),
\end{eqnarray}
where in the inertial frame ($S^\prime$) they arrive at different angular positions (Fig.2), that is,
\begin{eqnarray}\label{29}
\varphi^{\prime}_2-\varphi^{\prime}_1 = \Omega (t^{\prime}_2-t^{\prime}_1) = \Omega \Delta t^{\prime}.
\end{eqnarray}
So by substitution from (\ref{2600}) we have
\begin{eqnarray}\label{29-1}
{\Delta t} = {\Delta t}^{\prime}(\cosh\beta - \frac{R\Omega}{c}\sinh\beta),
\end{eqnarray}
which, in the Galilean limit $\beta = \frac{\Omega R}{c} \ll 1$, leads to,
\begin{eqnarray}\label{30}
{\Delta t}= \frac{4\pi R^2\Omega}{c^2}(1 + \frac{\beta^2}{2}  + {\cal O}(\beta^4)).
\end{eqnarray}
For the calculation of the same effect in the formalism based on Fermi metric we  use the relation between time coordinates in the inertial frame and the rotating one, namely Eq. (\ref{9}), from which the time interval between the two events $(t_1, y_1=0)$ and $(t_2, y_2=0)$ in the rotating frame is related to that in the inertial observer's frame as follows
\begin{eqnarray}\label{30-1}
{\Delta t} = \gamma^{-1}{{\Delta t}^{\prime}},
\end{eqnarray}
which is obviously the same as that obtained in \eqref{new2}. Substituting for the inertial time interval from (\ref{260}) and expanding in terms of $\beta$ we end up with,
\begin{eqnarray}\label{30-2}
{\Delta t} = 4\pi {R}^2 \frac{\Omega}{c^2}(1 + \frac{1}{2} \beta^2 + {\cal O}(\beta^4)).
\end{eqnarray}
Comparison of Eqs.  (\ref{30}) and (\ref{30-2}) shows that the two formalisms, as expected, agree up to the second order in $\beta$  but start to differ in the next order (fourth order in $\beta$). \\
For a possible experimental setup to measure this effect one could think of an observer sitting on the equator of a solid sphere (say Earth) rotating about its axis with angular velocity $\Omega$ which has a source and detector to send and receive corotating and counterrotating light beams. Such an observer would obviously be a non-inertial observer and the two approaches mentioned above should be employed in principle to predict the interference measurements. It should be noted that in the case of an observer in the equator of Earth, there would also be a general relativistic contribution, rooted in the axisymmetric nature of the spacetime metric, in this case that of Kerr weak field, produced by the sphere's (Earth's) rotational inertia, which leads to the so-called gravitomagnetic clock effect \cite{Coh}. This contribution being proportional to the mass of the source is different from the kinematical effect discussed above and in principle distinguishable from it \cite{Bosi}. Apart from the difficulty in measuring the fourth order difference, there is a subtlety in measuring the Sagnac effect in a non-inertial rotating observer's frame, however. This  arises from the fact that unlike the measurement of the transverse Doppler effect in which the angular velocity is an input (as in K{\"{u}}ndig experiment \cite{Kundig}), the Sagnac effect is mostly used as a rotation sensor (in the inertial observer's frame) to measure (and/or control) the angular velocity of the rotating platform on which the whole setup in mounted. This is the case, for example in both optical-fiber and laser ring interferometers used in navigation system gyroscopes or in rotational seismometers \cite{Gyro}. So to examine the above relation to the fourth order in $\beta$, one should try to design an experimental setup in which a secondary effect, originated from the time difference, such as the beat frequency in a ring laser interferometry (whose period is linearly proportional to the angular velocity of the platform) could be measured in both inertial and non-inertial frames of reference \cite{Post}.
\subsection{Length Measurement}
The length measurements by accelerated observers and their limitations in the formalism based on the hypothesis of locality (and Fermi coordinates) are 
discussed in \cite{Mashhoon2} and it is shown that the arclength, subtended by angle $\Phi$ between two uniformly rotating points on a circle of radius $R$ (such that $\beta^2 = \frac{R^2\Omega^2}{c^2} \ll 1$), as measured in the local frame of an observer (frame S) in one of those points is given by 
\begin{eqnarray}\label{31}
l= \frac{1}{\sqrt{1 - {\beta^2}}} [1-\frac{3}{4}\beta^2(1+\frac{\sin2\Phi-8\sin\Phi}{6\Phi})] l^\prime,
\end{eqnarray}
in which  $l^\prime= R\Phi$ is the same arclength in an inertial/laboratory observer's frame (frame $S^\prime$) who, in turn, assigns a contracted length to it as compared to the arclength $l^{\prime\prime}= \gamma l^\prime$ measured by a {\it comoving} inertial observer (frame $S^{\prime\prime}$).
Expansion to the second order in $\beta$ gives
\begin{eqnarray}\label{31-1}
l= [1 - \frac{1}{4}\beta^2 ( 1 + (\frac{\sin2\Phi-8\sin\Phi}{2\Phi})] l^\prime + {\cal O} (\beta^4),
\end{eqnarray}
which for small $\Phi$ (such that $\sin\Phi \approx \Phi$) reduces to
\begin{eqnarray}\label{31-2}
l= (1 + \frac{1}{2}\beta^2 ) l^\prime + {\cal O} (\beta^4),
\end{eqnarray}
corresponding to length dilation. But when the circumference ($L$) is found by setting $\Phi= 2\pi$, the observer arrives at the following relation in his local frame:
\begin{eqnarray}\label{31-3}
L = (1 - \frac{1}{4}\beta^2 ) 2\pi R + {\cal O} (\beta^4).
\end{eqnarray}
In other words small arclengths are dilated in the rotating frame but the whole path is contracted and this is so because for comoving inertial observers the circular path is momentarily an ellipse whose semi-minor axis is along the direction of the observer's motion (see Fig. 4 in \cite{Mashhoon2}). But it should be noted  that the same observer, attached to a rotating disk, finds his distance from the center of the disk to be always equal to the instantaneous ellipse's semi-major axis, which is equal to the disk radius $R$, and hence on returning to the same point (on the underlying spacetime) he finds out that he has moved on a circular path with circumference $2 \pi R$. This is consistent with the flat {\it spatial} geometry the observer assigns to the spacetime using Fermi metric on his worldline ($x^l=0$ in \eqref{4} or  $x=y=0$ in \eqref{10}) \cite{mtw}. This could be thought of as another manifestation of the so-called  Ehrenfest's paradox discussed in the literature \cite{MFT}.\\
Obviously in the above scenario one could take the two points separated by the arclength, to be on the rim of a uniformly rotating disk of radius $R$. With this setup one can use the formalism based on MFT to find the above discussed relation between the arclength measurements by  inertial and rotating observers, both in the Galilean limit. 
This can be obtained directly from the inverse angular transformation in (\ref{13}) by setting $\Delta t =0$ as follows:
\begin{eqnarray}\label{32}
l^\prime \equiv R d\phi^\prime =  \cosh (\beta) R d\phi \equiv \cosh (\beta) l,
\end{eqnarray}
in which one can think of $l$ as the small arclength subtended by the rotating observer's open feet.
Expanding the above relation to the second order in $\beta$ we end up with
\begin{eqnarray}\label{32-1}
l = (1 + \frac{1}{2}\beta^2 )l^\prime + {\cal O} (\beta^4),
\end{eqnarray}
which compared to (\ref{31-2}) shows that, to the second order in $\beta$, the two formalisms agree on the relation between the small arclengths as measured by the rotating (non-inertial) and inertial observers, and the difference shows up at the fourth order.\\ 
To calculate the circumference of the disk, the rotating observer finds the following relation between his measurement and that of the inertial observer:
\begin{eqnarray}\label{32-2}
L = (1 + \frac{1}{2}\beta^2 ) 2\pi R + {\cal O} (\beta^4).
\end{eqnarray}
In other words, contrary to the relation (\ref{31-3}), employing  MFT the rotating observer finds that the circumference of the path (disk) is dilated.\\
On the other hand, again on the observer's world line, i.e. at $r=R$,  the spacetime metric in \eqref{14} reduces to the Euclidean metric and consequently the circumference of the disk is found to be $2 \pi R$. In other words using the spacetime metric on the observer's world line, both formalisms give the same value $2 \pi R$ for the disk circumference but different values if the corresponding transformations are applied. This shows that the discussion of the length measurements, in both formalisms, is further complicated by the inclusion of Ehrenfest's paradox \cite{MFT}.
\section{Discussion}
In the present article first we introduced two fundamentally different approaches to study rotational phenomena as measured by non-inertial rotating  observers. The first approach employs consecutive Lorentz transformation on the basis of the so-called hypothesis of locality in which an accelerated observer (frame) is taken to be instantaneously equivalent to inertial observers (frames) which have the velocity of the accelerated frame at each moment on its worldline. Applying it to the special case of rotational acceleration (e.g. on a rigidly rotating disk) one arrives at the coordinate transformations
given by (\ref{9}). This is, in essence, special relativity (and local Lorentz transformations) applied to rotating frames in which the velocity distribution is given by the classical relation $v=r\Omega$. In the second approach a kinematical  transformation  between inertial  and rotating frames was given by introducing a {\it relativistic rotational transformation}  called MFT, which is intrinsically different from the Lorentz transformation which only applies to inertial frames at uniform relative motion. The main characteristic of this approach is the introduction of a nonlinear velocity distribution  $v=c\tanh \frac{r\Omega}{c}$ on a uniformly rotating platform. \\
Two well known rotational effects, transverse Doppler effect and Sagnac effect, were studied in the Galilean limit ($\beta  \ll 1$) in both formalisms to find how their predictions on the relation between the quantities measured by inertial and rotating observers differ from one another. It should be noted that the rotational effects discussed are not just the artifact of coordinates and their differences, since in both formalism a unique setup was analyzed in which a non-inertial observer at nonzero radius on a uniformly rotating disk is involved in the measurements.\\
In both effects the difference in the predicted relation between the quantities measured by inertial on rotating observers starts at the fourth order in $\beta$. The agreement between the predictions in the two formalisms, up to the second order in $\beta$, was expected from the fact that the time-time components of the corresponding metrics (in the rotating frame) in both formalisms were equal up to the same order in $\beta$. Indeed since our results and those of Mashhoon et al. agree  up to the second order in $\beta$, the precision of the experiments carried out on rotating platforms are still far from showing deviations which could differentiate between the two formalisms.\\
Also the lifetime (energy) of an orbiting unstable particle fixed at nonzero radius on a rigidly rotating disk, measured by a comoving  eccentric observer, and its lifetime (energy) measured in the inertial observer's frame, could be related through MFT. Again, up to the second order in $\beta =R\Omega/c$,  it is found to be in agreement with the relation obtained through the application of instantaneous Lorentz transformation \cite{MFT}. But one should be cautious and differentiate between the experiments carried out on rotating platforms (e.g. a rigidly rotating disk), and those on forced circular orbits such as in the CERN Muon storage ring in which the lifetimes of muons have been measured \cite{Bailey}.\\
In the measurements of small arclengths, again the two formalisms agree up to the second order in $\beta$ and differ at the fourth order and higher. The two approaches disagree on the relation between the circumference of the circular path as measured by an inertial observer and a rotating one at nonzero radius. Measurements of the rotating observer compared to the inertial one imply contraction of the circumference in the formalism based on  Fermi metric  and  dilation in the formalism based on MFT. All this is true if one avoids discussing Ehrenfest's paradox, by only working with the coordinate transformations and not their corresponding spatial metrics (geometries). If on the other hand one uses spacetime metric on the observer's world line, the two approaches lead to the same value $2 \pi R$ for the circumference of the disk. In other words, once again we face Ehrenfest's paradox which further complicates the discussion of the length measurements in both formalisms \cite{MFT}.
\section *{Acknowledgments}
The authors would like to thank University of Tehran for supporting this project under the grants provided by the research council.
\pagebreak

\end{document}